\documentclass[sigconf]{acmart}
\usepackage[acronym]{glossaries}

\usepackage{soul,color}
\usepackage{siunitx}
\usepackage{multirow}
\usepackage{makecell}
\usepackage{pdflscape}
\usepackage{tabularx,array}
\usepackage{tikz}
\usepackage{amsfonts}

\usetikzlibrary{arrows.meta,positioning,shapes.geometric,fit}

\newcolumntype{L}{>{\raggedright\arraybackslash}X}
\newcolumntype{Y}{>{\centering\arraybackslash}X}
\newcolumntype{Z}{S[table-format=1.3]}

\newacronym{tar}{TAR}{Technology Assisted Review}
\newacronym{dta}{DTA}{Diagnostic Test Accuracy}
\newacronym{fpr}{FPR}{False Positive Rate}
\newacronym{pmid}{PMID}{PubMed Identifier}

\DeclareMathOperator*{\argmin}{arg\,min}

\AtBeginDocument{%
  }

\copyrightyear{2026}
\acmYear{2026}
\setcopyright{cc}
\setcctype{by}
\acmConference[SIGIR '26]{Proceedings of the 49th International ACM SIGIR Conference on Research and Development in Information Retrieval}{July 20--24, 2026}{Melbourne, VIC, Australia}
\acmBooktitle{Proceedings of the 49th International ACM SIGIR Conference on Research and Development in Information Retrieval (SIGIR '26), July 20--24, 2026, Melbourne, VIC, Australia}
\acmDOI{10.1145/3805712.3809924}
\acmISBN{979-8-4007-2599-9/2026/07}

\begin{document}

\title{Confidence-Based Stopping Methods for Systematic Reviews}
\author{Aaron H.A. Fletcher}
\orcid{0000-0002-4776-066X}
\affiliation{%
  \department{School of Computer Science}
  \institution{University of Sheffield}
  \city{Sheffield}
  \country{United Kingdom}
}
\email{ahafletcher1@sheffield.ac.uk}

\author{Mark Stevenson}
\orcid{0000-0002-9483-6006}
\affiliation{%
  \department{School of Computer Science}
  \institution{University of Sheffield}
  \city{Sheffield}
  \country{United Kingdom}
}
\email{mark.stevenson@sheffield.ac.uk}

\begin{abstract}
Technology Assisted Review stopping methods aim to ensure that no more documents are screened than necessary. Most existing approaches focus on achieving a target recall, which does not consider whether an information need has been met. This paper introduces two heuristic stopping methods that instead monitor whether screened documents contain enough information to make a decision. Evaluation on a standard dataset of Diagnostic Test Accuracy Systematic Reviews demonstrates that the proposed approaches substantially reduce the number of documents that need to be examined while, in the majority of cases, maintaining conclusions that are consistent with all evidence available. 
\end{abstract}



\renewcommand{\shortauthors}{Aaron H.A. Fletcher and Mark Stevenson}
\maketitle

\section{Introduction}

\Gls{tar} develops methods to identify information when it would be impractical to examine an entire collection \cite{nunzio_special_2023}. Key applications include the development of systematic reviews \cite{higgins2019cochrane} and eDiscovery \cite{roegiest2015trec,grossman2016trec,oard2018jointly,mcdonald2020accuracy}. Within \Gls{tar}, stopping rules help reviewers determine when to stop examining documents, thereby reducing the workload by reviewing only the necessary documents. A wide range of stopping methods have been proposed, most of which base the decision to stop on having identified a desired portion of all the relevant documents within the collection, known as the \emph{target recall}~\cite{li2020stop,stevenson2023stopping}. However, basing the stopping decision on target recall does not consider whether a user's information need has been satisfied. Under certain circumstances, a user's information need may be satisfied with information contained within a single document, even when many other relevant documents in the collection have not been identified. In other cases, the information need cannot be satisfied even after screening the entire collection.

An alternative approach is to consider whether the user's information need has been satisfied, rather than whether target recall has been achieved. This approach is formalised in this work within the context of developing Systematic Reviews, specifically those that determine the accuracy of diagnostic tests, such as blood tests for viral infection. The accuracy of such tests is reported in \Gls{dta} studies using clinical metrics such as \Gls{fpr} and Sensitivity.\footnote{The sensitivity metric is more commonly known as recall within Information Retrieval. We use the term ``sensitivity'' to refer to estimates of diagnostic test accuracy reported within scientific papers while reserving the term ``recall'' to its standard usage, i.e. proportion of relevant documents identified.} However, the values reported in individual studies vary due to factors such as different sample sizes and populations tested. DTA Systematic Reviews aim to summarise all available data to produce the best available estimate of test accuracy. However, systematic reviews are expensive and time-consuming to carry out, making them impractical in many scenarios. Instead, an individual may want to understand whether the test is accurate enough for some purpose or not. This paper introduces two stopping algorithms for this scenario, each designed to determine whether the examined evidence is sufficient to answer the information need. The approaches monitor the evolving point estimates and uncertainty of the clinical metrics and stop when the accumulated evidence appears sufficient to support the thresholded decision.

These approaches are compared against existing stopping rules (most of which base their decisions on target recall) and found to improve the efficiency of the review process. They achieve comparable results to target recall-based methods while examining substantially fewer documents.  

This paper contributes by: (i) introducing confidence-based stopping rules that focus on determining when an information need has been met, rather than when a target recall has been achieved; (ii) developing two such rules for a systematic review setting; (iii) evaluating those rules on a real-world dataset of \Gls{dta} reviews.

\section{Background}

\Gls{tar} stopping rules aim to cut screening effort while meeting an acceptability constraint \cite{cormack2016engineering,li2020stop,lewis2021certifying,stevenson2023stopping}. Most methods monitor progress toward a \emph{target recall}, e.g. 70\% of all relevant documents in the collection, and stop once enough relevant documents have been observed that the target is likely to have been met or even exceeded. This recall-focused approach has historical roots in legal information retrieval \cite{blair1985evaluation}.

Approaches include observing the yield of relevant items over time \cite{cormack2016engineering,stevenson2023stopping}, estimating remaining relevant items via sampling or classification \cite{shemilt2014pinpointing,callaghan2020statistical,cormack2016scalability,li2020stop,molinari2024saltau, bron2025using}, analysing rank scores for inflection points \cite{hollmann2017ranking,di2018study} and reinforcement learning ~\cite{bin-hezam_rlstop_2024, Bin_Hezam_2025}. Evaluation typically examines whether the method reliably attains the target recall \cite{cormack2016engineering} or how far achieved recall deviates from the target \cite{li2020stop,stevenson2023stopping}. Framing stopping around recall is attractive because it rests on the familiar notion of relevance, is straightforward to implement with standard test collections (e.g., CLEF e-Health \cite{kanoulas2017clef,kanoulas2018clef,kanoulas2019clef}, TREC Total Recall \cite{grossman2016trec}, TREC Legal \cite{cormack2010overview}, RCV1 \cite{lewis2004rcv1}), and aligns with eDiscovery where parties are required to achieve a target recall \cite{yang2021minimizing}.

However, recall assesses the proportion of relevant documents discovered and not whether the user's information need has been met. It also treats all relevant items as equally valuable \cite{saracevic1988study,saracevic2022notion} which is not the case in systematic reviews where studies contribute unequally to conclusions; large, high-quality studies often dominate the evidence and those reporting experiments carried out over small samples add little marginal value. It has been demonstrated that conclusions may be clear without the need to examine all relevant documents \cite{norman2019measuring}. Consequently, a recall threshold can force unnecessary screening even when additional evidence does not change the answer.

The present work differs from previous TAR stopping methods in that it does not target a recall threshold, but instead monitors whether the accumulated evidence is sufficient for a downstream decision.

Our approach has some parallels with work on the analysis of stopping behaviour in interactive search which demonstrated that individuals stop searching when information feels ``good enough,'' balancing benefits and costs under cognitive and environmental constraints \cite{ilani2024analysis,cooper1973selecting,prabha2007enough,zach2005enough,dostert2009users,wu2014online,pennington2016much,azzopardi2013query} and economic models that treat stopping interactive search as an economic problem by maximising expected utility net of search costs \cite{azzopardi2011economics}. However, our focus is on determining when to stop examining ranked lists of documents rather than interactive search.

\section{Stopping Methods}

Assume that a collection of documents has been ranked and that its documents are examined in that order. Let $i$ be the $i$th relevant study encountered in the ranking. When retrieving documents, a point estimate, $\hat{p}_i$, for a metric of interest (e.g., sensitivity or \Gls{fpr}) can be derived from the cumulative information gathered so far. We further assume that the user wishes to know whether the value of this metric is greater or less than some threshold, $\theta$. The goal is to find the smallest index, $I^*$, such that sufficient information has been acquired for the information need to have been met. 

The methods proposed here instantiate a confidence-based perspective through two heuristic rules which monitor the evolving value of $\hat{p}_i$ and its uncertainty. 

\noindent{\bf Harmonic Shrinkage}. The first approach sets a boundary around the value of $\theta$ which is gradually narrowed as more estimates of $\hat{p}_i$ become available. The approach stops when all values of $\hat{p}_i$ lie outside this boundary, either all above or below it. The approach is formalised as: 
\begin{equation}
  \label{Harmonic Shrinkage Based Stopping}
  \begin{aligned}
  I^* = \argmin_{i}  \left(\left( \forall j \leq i, \; \hat{p}_j \ge \theta  + \frac{s (1-\theta)}{i} \right)  \lor \left( \forall j \leq i, \;  \hat{p}_j \le \theta - \frac{s \theta}{i} \right)\right)
  \end{aligned}
  \end{equation}

where $s$ is a hyperparameter that controls the convergence rate of the decision boundary.

This approach will never stop if the estimates straddle the decision threshold for any pair of estimates, i.e. $\hat{p_k} > \theta$ and $\hat{p_l} < \theta$ for some $k, l \leq i$, potentially limiting its applicability when the true value is close to the decision threshold.  
  
\noindent{\bf Confidence Boundary}. Another approach is to construct confidence intervals around each $\hat{p}_i$ using estimated variance. 

This approach addresses a limitation of the \textsc{harmonic shrinkage} method, namely its sole reliance on point estimates and its failure to consider the variability of $\hat{p}_i$. Let $MOE_{0.95}$ be the margin of error corresponding to the 95\% confidence interval calculated from the cumulative data observed so far. The \textsc{Confidence Boundary} approach is defined as follows: 
\begin{equation}
\label{Confidence Interval}
I^* = \argmin_{i} \left(  \left(\hat{p}_i + MOE_{0.95} < \theta \right) \lor \left(\hat{p}_i - MOE_{0.95} > \theta \right) \right)
\end{equation}

In other words, this approach stops when the estimate is at least one margin of error away from $\theta$. 

\section{Experiments}
Experiments evaluate the performance of the proposed stopping methods along two dimensions: their ability to reach the correct conclusion given the evidence while also minimising the number of documents that need to be examined.

\subsection{Dataset}

Evaluation is based on the \Gls{dta} systematic reviews from the Cochrane Library available from the CLEF Technology-Assisted Review in Empirical Medicine exercises (CLEF 2017/2018/2019) \cite{kanoulas2017clef,kanoulas2018clef,kanoulas2019clef} which have been widely used in previous work on TAR stopping. These data sets consist of scientific articles from the PubMed database, identified uniquely by a \Gls{pmid}, labelled for relevance depending on whether or not they contain information about a diagnostic test's accuracy. 

Each review addresses multiple {\it outcomes}, each corresponding to an evaluation of the diagnostic test within a specific application (e.g. to a particular age group), with relevant \Gls{pmid}s containing information about one or more of the review's outcomes. In order for the data set to be useful to evaluate the approaches proposed in this paper we need information about the test accuracy for each outcome reported in each relevant \Gls{pmid}. Fortunately much of this is available in the LIMSI-Cochrane dataset \cite{norman2018data} which contains diagnostic counts of test accuracy (i.e. True Positives, True Negatives, False Positives and False Negatives) for relevant documents in CLEF 2017 and 2018. 
 
The LIMSI-Cochrane dataset was constructed using a combination of optical character recognition technology, manual verification and post-editing with a reported low extraction error rate (0.06–0.3\%). Since LIMSI-Cochrane predates the CLEF 2019 reviews, the required information for studies within these reviews were extracted directly from source documents available on the Cochrane Library website using AWS Textract OCR technology and linked to the corresponding \Gls{pmid}. In some cases, it was not possible to extract diagnostic counts, and such \Gls{pmid}s were treated as being irrelevant (i.e. do not contain useful information for determining the outcome).

The evaluation dataset was constructed by considering each outcome individually. The set of \Gls{pmid}s from which diagnostic counts for each outcome were available was identified and the outcome retained only if there were at least five such \Gls{pmid}s. This produced a data set consisting of systematic review outcomes for which \Gls{pmid}s identified as relevant also included diagnostic counts for that outcome. Because each systematic review may address multiple outcomes, the number of relevant \Gls{pmid}s for an individual outcome is typically smaller than the number of relevant \Gls{pmid}s for the full review. This procedure yielded a dataset containing 135 qualifying outcomes: 90 from CLEF 2017 reviews, 29 from CLEF 2018, and 16 from CLEF 2019.

For our evaluation, we also require a full-evidence reference value for the sensitivity and \Gls{fpr} for each outcome. A systematic review contains the best estimate of these values using all evidence available at the time of its creation so we use the value produced by considering all extracted relevant studies for an outcome as the full-evidence reference value. This may differ from the value reported in the original review if diagnostic counts could not be extracted from some \Gls{pmid}s.

\subsection{Implementation}

The complete set of documents for each outcome is ranked by AutoTAR \cite{cormack2015autonomyreliabilitycontinuousactive} using a reference implementation \cite{li2020stop} with an initial sample of 100 documents. For each outcome, documents are then examined in ranked order and when each relevant document is encountered the sensitivity and \Gls{fpr} estimates are updated using the Reitsma bivariate meta-analytic model estimate~\cite{reitsma_bivariate_2005} to produce the sequence of estimates $\{\hat p_i\}$ associated with each relevant document. Reitsma also computes the confidence interval ($MOE_{0.95}$) used by the \textsc{Confidence Boundary} approach. 

Decision thresholds ($\theta$) were set to 0.75 for sensitivity and 0.1 for \Gls{fpr}. These values were selected because of their clinical plausibility, e.g. a test with a sensitivity of 0.75 is potentially useful if it is inexpensive and non-invasive compared to alternative diagnostic procedures, despite missing a large proportion of the population with a disease.  

For \textsc{Harmonic Shrinkage}, the convergence rate, $s$, was set to 0.25 to balance early stopping and decision stability. 

\subsection{Baselines}

The proposed stopping algorithms were compared against several baselines. Several of these approaches rely on hyperparameters which can be used to create conservative and aggressive versions which bias it towards later and earlier stopping, respectively. The conservative variants are Knee (-), Target (5), and SCAL (0.9), while Knee (2), Target (3), and SCAL (1.05) represent more aggressive variants.

\noindent \emph{70\% Recall Baseline:} Stop after at least 70\% of relevant studies are retrieved. This represents a common heuristic used in some \Gls{tar} applications, aiming to balance search effort with the goal of finding a substantial portion of relevant material. This baseline relies on perfect knowledge of the relevant documents in the collection, which is typically unknown during retrieval.

\noindent \emph{Knee:} Monitor the recall-versus-cost (gain curve) during the review process and stop when the rate of finding new utility-labelled studies significantly decreases relative to the initial rate, identified algorithmically as the `knee' point of diminishing returns on the curve \cite{cormack2016engineering}. Two variants are used with the $\rho$ parameter set to 2 and ``dynamic'': ``Knee (2)'' and ``Knee (-)''.

\noindent \emph{Target:} Randomly samples a set number of relevant studies, creating a target set. Then, a standard \Gls{tar} process runs until all documents in this initial target set have been reviewed \cite{cormack2016engineering}. 
The target set value, $|T|$,  is set to 3 and 5, indicated by ``Target (3)'' and ``Target (5)''.

\noindent \emph{SCAL (Scalability of Continuous Active Learning):} Iteratively train rankers on a large random sample from the collection, then estimate the total number of relevant studies ($R$) via stratified sub-sampling (calibrated for the full collection) to determine a score threshold. This threshold, which achieves a target recall based on $R$, is applied to the final ranked list of the entire collection to select documents \cite{cormack2016scalability}. SCAL's target recall is set to 0.7 and the $\eta$ values to 1.05 and 0.9: ``SCAL (1.05)'' and ``SCAL (0.9)''.

With the exception of \textsc{70\% Recall}, all baselines were computed using the reference implementations described by~\citet{li2020stop}.

\begin{table*}[t!]

    \centering
 
    \caption{Performance of confidence-based and baseline stopping algorithms across 135 DTA review outcomes, evaluated on False Positive Rate and Sensitivity. Bold indicates best DA, Sav., and AS. Stops and Rec. are not highlighted as optimal, as they are task-dependent.} 

    \begin{tabular}{|l@{\hspace{0.5em}}|lllll|lllll|}
    \hline
     & \multicolumn{5}{c|}{\textit{False Positive Rate}} & \multicolumn{5}{c|}{\textit{Sensitivity}}\\
     
     & \textbf{DA} & \textbf{Sav.} & \textbf{Stops} & \textbf{AS}  &  \textbf{Rec.} & \textbf{DA} & \textbf{Sav.} & \textbf{Stops} & \textbf{AS} & \textbf{Rec.}\\
    \hline
 \multicolumn{11}{|l|}{\textit{Baselines}} \\
 \hline
 70\% Recall & 0.963 &  0.900 & 1.000 & 0.719 & 0.751 & 0.889  & 0.900 & 1.000 & 0.578 & 0.751\\
    Knee (2) & 0.993 & 0.747 & 0.993 & 0.726 & 0.909 & 0.981  & 0.747 & 0.993 & 0.585 & 0.909\\
    Knee (-) & \textbf{1.000} &  0.079 & 0.169 & 0.382 & 1.000 & \textbf{1.000}  & 0.079 & 0.169 & 0.487 & 1.000\\
    SCAL (1.05) & \textbf{1.000} &  0.600 & 0.959 & 0.695 & 0.983 & \textbf{1.000}  & 0.600 & 0.959 & 0.578 & 0.983\\
    SCAL (0.9) & \textbf{1.000} &  0.560 & 0.821 & 0.584 & 0.982 & \textbf{1.000}  & 0.560 & 0.821 & 0.585 & 0.982\\
    Target (3) & \textbf{1.000} &  0.486 & 0.905 & 0.647 & 0.999 & 0.996  & 0.486 & 0.905 & 0.578 & 0.999\\
    Target (5) & \textbf{1.000} &  0.332 & 0.916 & 0.679 & 0.999 & 0.997  & 0.332 & 0.916 & 0.547 & 0.999\\
    \hline
    \multicolumn{11}{|l|}{\textit{Confidence-based}} \\
    \hline
    \textsc{Harmonic Shrinkage} & 0.889 &  \textbf{0.956} & 1.000  & 0.719 & 0.180  & 0.807 &  \textbf{0.933} & 0.978 & 0.600 & 0.175\\
    \textsc{Confidence Boundary} & 0.970 &  0.853 & 0.896 & \textbf{0.822} & 0.309 & 0.926  & 0.741 & 0.778 & \textbf{0.800} & 0.415\\
    \hline
    \end{tabular}
    \label{tab:stopping-comparison-combined}
\end{table*}

\subsection{Evaluation metrics}

\noindent{\bf Decision Agreement (DA)}. Following Kusa et al.~\cite{kusa2023outcome}, a stopping algorithm's effectiveness was evaluated by calculating the proportion of outcomes for which the decision when the algorithm stops is the same as the decision would have been given all potential evidence. 

The decision at index $i$ is a binary value given by 
\begin{equation}
\label{decision}
\text{Decision}(i) =
\begin{cases}
1, & \text{if } (\hat{p}_i > \theta \text{ for Sens.}) \ \lor\ (\hat{p}_i < \theta \text{ for FPR})\\
0, & \text{otherwise}
\end{cases}
\end{equation}

Then the decision agreement for a set of outcomes, $O$, is calculated as the proportion of times the decisions at the stopping point ($Decision(I^{*})$) and when all information is considered ($Decision(n)$) agree over all outcomes, i.e., 
\begin{equation}
\label{DA}
\text{Decision Agreement}
=
\frac{1}{|O|}
\sum_{o \in O}
\mathbb{I}\!\left(
\operatorname{Decision}_o(I_o^{*})
=
\operatorname{Decision}_o(n_o)
\right).
\end{equation}

where $\mathbb{I}$ is the indicator function and $n_o$ denotes the index after all relevant studies for outcome $o$ have been retrieved.


\noindent{\bf Savings (Sav.)} The percentage of documents in the collection not reviewed before the stopping point, averaged over all outcomes.

\noindent{\bf Stops} The percentage of outcomes where the stopping algorithm is activated, i.e. does not examine all documents in the ranking. 

\noindent{\bf Appropriate Stops ({\bf AS})} Stopping is not always appropriate, such as when the collection lacks sufficient information to address the information need. A stopping decision is deemed appropriate if either: (1) the algorithm stopped when sufficient information was available for a confident decision, or (2) the algorithm continued when insufficient information was available. Sufficient information is defined as a state where the decision threshold ($\theta$) lies outside the margin of error of the point estimate \emph{after} all relevant studies have been retrieved. 

Note that Appropriate Stops reflects alignment with the statistical sufficiency of the full evidence, not a cost-sensitive utility model. The evaluation, therefore, assesses \emph{decision consistency under partial evidence}.

Let $S_o$ be a binary variable indicating whether the algorithm stopped for outcome $o$ (1 = stopped, 0 = did not stop) and $I_o$ be a binary variable indicating whether information was sufficient for outcome $o$ (1 = sufficient, 0 = insufficient). Appropriate stopping is then calculated as the percentage of outcomes where the stopping decision was appropriate: 
\begin{equation}
\label{AppStops}
\text{Appropriate Stops}
=
\frac{1}{|O|}
\sum_{o \in O}
\mathbb{I}(S_o = I_o).
\end{equation}

\noindent{\bf Recall ({\bf Rec.})} The percentage of {\it all relevant} documents in the collection reviewed before the stopping point, also averaged over all outcomes. Recall remains an informative measure, although, unlike in many other Information Retrieval scenarios, our goal here is not to maximise it. 

\section{Results}

Table \ref{tab:stopping-comparison-combined} presents the aggregated performance of the proposed confidence-based stopping algorithms compared against baseline methods across 135 \Gls{dta} review outcomes. Results demonstrate that the proposed approaches, particularly \textsc{Confidence Boundary}, substantially reduce the number of documents that need to be examined while largely preserving decision agreement.

Savings for the majority of baselines (excluding \textsc{70\% Recall} which cannot be deployed in practice) are lower than those of the proposed approaches, demonstrating that baselines require more of the collection to be examined prior to stopping screening. Several of the baselines (\textsc{Knee (-)}, \textsc{Target (3)} and \textsc{Target (5)}) require more than half of the collection to be examined while the proposed approaches reduce this to around a quarter, or less. These savings are reflected in the Recall scores which are noticeably lower for the proposed approaches than the baselines which typically retrieve almost all relevant documents before stopping, which the Decision Agreement scores demonstrate is not necessary.

Efficiency gains were achieved while largely maintaining high effectiveness. Although some baselines achieved perfect DA, the top-performing of the proposed approaches also yielded high DA scores. \textsc{Confidence Boundary} consistently achieved high agreement (DA: 97\% FPR, 92.6\% Sens).

\textsc{Confidence Boundary} also proved able to make appropriate stopping decisions, outperforming all other methods on the AS metric. This highlights the core benefit of aligning the stopping decision with statistical sufficiency of the accumulated evidence rather than recall targets, which are agnostic to whether the review question can yet be answered reliably. In contrast, baseline methods exhibited lower AS scores ($\leq$72.6\%), precisely because their stopping criteria (e.g., target recall) are independent of final information sufficiency.

\begin{figure}
    \centering
    \includegraphics[width=0.95\linewidth]{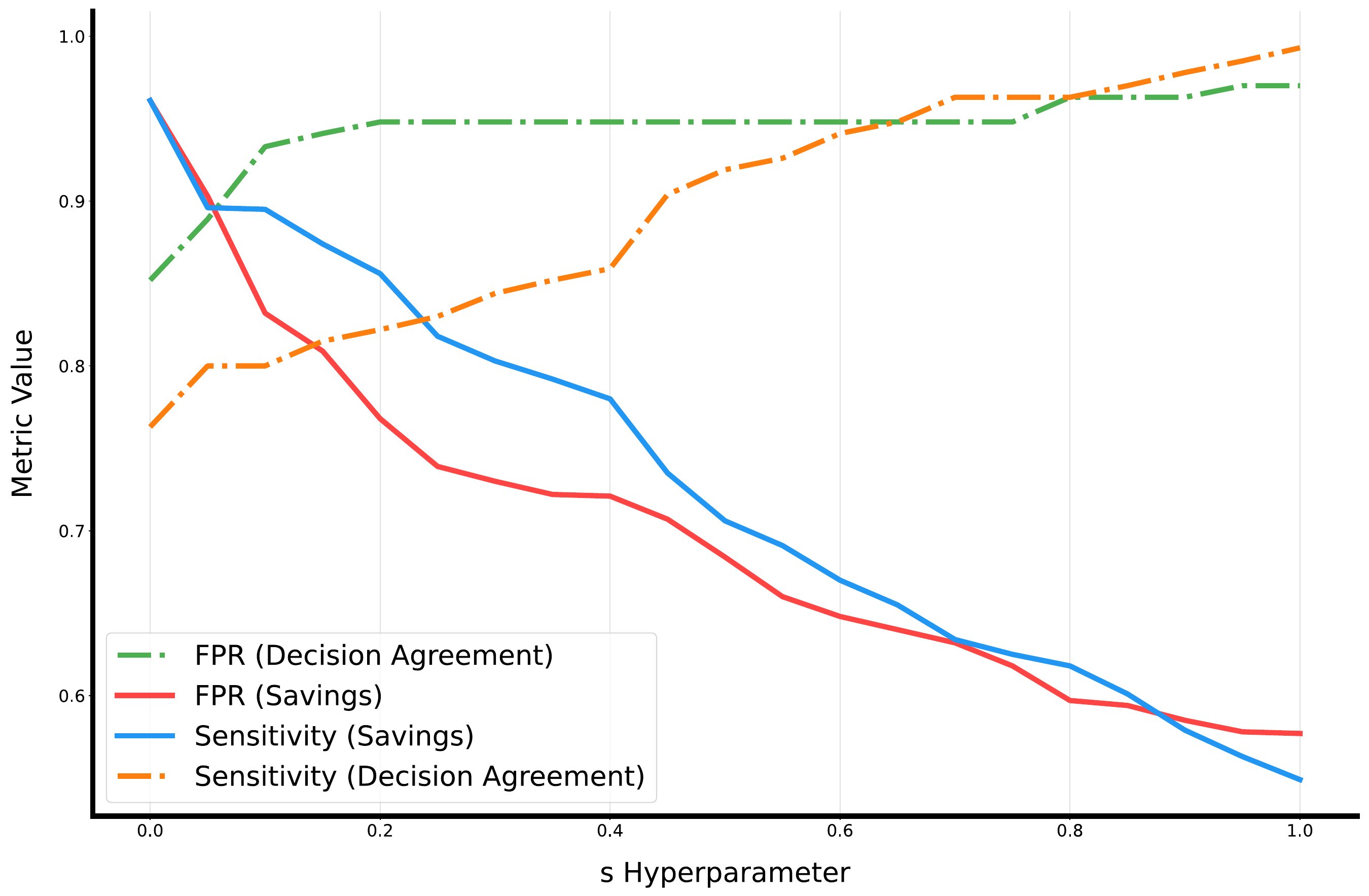}
    \caption{Effect of changing the $s$ hyperparameter for \textsc{Harmonic Shrinkage}. Increasing $s$ results in greater DA, at the cost of savings.}
    \Description{A line plot showing the effect of the harmonic shrinkage hyperparameter s on decision agreement and savings. As s increases, decision agreement rises while savings fall.}
    \label{fig:hyperparam_s}
\end{figure}

Varying the value of hyperparameter $s$ in the \textsc{Harmonic Shrinkage} approach had the expected effect: increasing $s$ resulted in greater decision agreement but reduced savings, and vice versa, as shown in Figure \ref{fig:hyperparam_s}.

The stopping methods proposed here may struggle with small sample sizes and skewed data distributions, which are common challenges in systematic reviews, particularly for those addressing rare diseases or conditions. With few data points, estimates become unstable and sensitive to individual studies, potentially leading to premature or delayed stopping. 

The approaches also assume that the sampling distribution of the point estimates is approximately normal. While the Central Limit Theorem suggests that this assumption holds for sufficiently large samples, it may not for small sample sizes. 

\section{Conclusion}

This paper has introduced an alternative approach to \Gls{tar} stopping that focuses on whether retrieved information is sufficient to support a downstream decision rather than on whether a predefined recall threshold has been reached. Experimental results from systematic \Gls{dta} reviews demonstrate that the proposed approaches can greatly reduce relevant document screening requirements while preserving decision agreement. By monitoring evolving point estimates and uncertainty for key clinical metrics, these approaches stop when the accumulated evidence appears sufficient for the thresholded decision. Overall, the results suggest benefits in shifting from recall-oriented stopping methods to confidence-based approaches for stopping decisions. 

The underlying principle, that stopping should be governed by evidence sufficiency for a thresholded decision rather than by recall targets, may transfer to other Information Retrieval settings, but the current study evaluates the methods exclusively on DTA systematic reviews. Extending the evaluation to intervention reviews, where outcome structure differs, and to non-medical TAR domains remains important future work.

Future work could extend this evidence-sufficiency framing into a more formal decision-theoretic approach, allowing practitioners to specify task-specific costs, benefits, and stopping criteria aligned with their information needs.

\begin{acks}
This work was supported by the UKRI AI Centre for Doctoral Training in Speech and Language Technologies (SLT) and their Applications funded by UK Research and Innovation [grant number EP/S023062/1].
\end{acks}

\bibliographystyle{ACM-Reference-Format}
\balance
\bibliography{sample-base}

\end{document}